\documentclass[prl,twocolumn,superscriptaddress,showpacs,amssymb,amsmath,amsfonts,aps]{revtex4}
\usepackage[dvips]{graphicx}
\usepackage{xspace}

\def\gpcm2{\ensuremath{\rm{g/cm^{2}}}\xspace}

\def\pair{\ensuremath{e^{+}e^{-}}\xspace}

\def\Kpair{\ensuremath{K^{+}K^{-}}\xspace}

\def\hnuc{\ensuremath{\rm{^{1}H}}\xspace}
\def\dnuc{\ensuremath{\rm{^{2}H}}\xspace}
\def\he3nuc{\ensuremath{\rm{^{3}He}}\xspace}
\def\car12{\ensuremath{\rm{^{12}C}}\xspace}
\def\ti48{\ensuremath{\rm{^{48}Ti}}\xspace}
\def\fe56{\ensuremath{\rm{^{56}Fe}}\xspace}
\def\cu63{\ensuremath{\rm{^{63}Cu}}\xspace}
\def\pb208{\ensuremath{\rm{^{208}Pb}}\xspace}
\def\linuc{\ensuremath{\rm{Li}}\xspace}
\def\cnuc{\ensuremath{\rm{C}}\xspace}
\def\alnuc{\ensuremath{\rm{Al}}\xspace}
\def\tinuc{\ensuremath{\rm{Ti}}\xspace}
\def\fenuc{\ensuremath{\rm{Fe}}\xspace}
\def\cunuc{\ensuremath{\rm{Cu}}\xspace}
\def\nbnuc{\ensuremath{\rm{Nb}}\xspace}
\def\pbnuc{\ensuremath{\rm{Pb}}\xspace}
\def\tife{\ensuremath{\rm{\fenuc-\tinuc}}\xspace}

\begin{document}

\preprint{Phys. Rev. Lett.}

\title{{\large Absorption of the $\omega$ and $\phi$ Mesons in Nuclei}}


\newcommand*{\CANISIUS}{Canisius College, Buffalo, NY 14208, USA}
\newcommand*{\CANISIUSindex}{1}
\affiliation{\CANISIUS}
\newcommand*{\GWU}{The George Washington University, Washington, DC 20052, USA}
\newcommand*{\GWUindex}{2}
\affiliation{\GWU}
\newcommand*{\SCAROLINA}{University of South Carolina, Columbia, South Carolina 29208, USA}
\newcommand*{\SCAROLINAindex}{3}
\affiliation{\SCAROLINA}
\newcommand*{\JLAB}{Thomas Jefferson National Accelerator Facility, Newport News, Virginia 23606, USA}
\newcommand*{\JLABindex}{4}
\affiliation{\JLAB}
\newcommand*{\ANL}{Argonne National Laboratory, Argonne, Illinois 60441, USA}
\newcommand*{\ANLindex}{5}
\affiliation{\ANL}
\newcommand*{\ASU}{Arizona State University, Tempe, Arizona 85287-1504, USA}
\newcommand*{\ASUindex}{6}
\affiliation{\ASU}
\newcommand*{\UCLA}{University of California at Los Angeles, Los Angeles, California  90095-1547, USA}
\newcommand*{\UCLAindex}{7}
\affiliation{\UCLA}
\newcommand*{\CSUDH}{California State University, Dominguez Hills, Carson, CA 90747, USA}
\newcommand*{\CSUDHindex}{8}
\affiliation{\CSUDH}
\newcommand*{\CMU}{Carnegie Mellon University, Pittsburgh, Pennsylvania 15213, USA}
\newcommand*{\CMUindex}{9}
\affiliation{\CMU}
\newcommand*{\CUA}{Catholic University of America, Washington, D.C. 20064, USA}
\newcommand*{\CUAindex}{10}
\affiliation{\CUA}
\newcommand*{\SACLAY}{CEA, Centre de Saclay, Irfu/Service de Physique Nucl\'eaire, 91191 Gif-sur-Yvette, France}
\newcommand*{\SACLAYindex}{11}
\affiliation{\SACLAY}
\newcommand*{\CNU}{Christopher Newport University, Newport News, Virginia 23606, USA}
\newcommand*{\CNUindex}{12}
\affiliation{\CNU}
\newcommand*{\UCONN}{University of Connecticut, Storrs, Connecticut 06269, USA}
\newcommand*{\UCONNindex}{13}
\affiliation{\UCONN}
\newcommand*{\EDINBURGH}{Edinburgh University, Edinburgh EH9 3JZ, United Kingdom}
\newcommand*{\EDINBURGHindex}{14}
\affiliation{\EDINBURGH}
\newcommand*{\FU}{Fairfield University, Fairfield CT 06824, USA}
\newcommand*{\FUindex}{15}
\affiliation{\FU}
\newcommand*{\FIU}{Florida International University, Miami, Florida 33199, USA}
\newcommand*{\FIUindex}{16}
\affiliation{\FIU}
\newcommand*{\FSU}{Florida State University, Tallahassee, Florida 32306, USA}
\newcommand*{\FSUindex}{17}
\affiliation{\FSU}
\newcommand*{\ISU}{Idaho State University, Pocatello, Idaho 83209, USA}
\newcommand*{\ISUindex}{18}
\affiliation{\ISU}
\newcommand*{\INFNFR}{INFN, Laboratori Nazionali di Frascati, 00044 Frascati, Italy}
\newcommand*{\INFNFRindex}{19}
\affiliation{\INFNFR}
\newcommand*{\INFNGE}{INFN, Sezione di Genova, 16146 Genova, Italy}
\newcommand*{\INFNGEindex}{20}
\affiliation{\INFNGE}
\newcommand*{\INFNRO}{INFN, Sezione di Roma Tor Vergata, 00133 Rome, Italy}
\newcommand*{\INFNROindex}{21}
\affiliation{\INFNRO}
\newcommand*{\ORSAY}{Institut de Physique Nucl\'eaire ORSAY, Orsay, France}
\newcommand*{\ORSAYindex}{22}
\affiliation{\ORSAY}
\newcommand*{\ITEP}{Institute of Theoretical and Experimental Physics, Moscow, 117259, Russia}
\newcommand*{\ITEPindex}{23}
\affiliation{\ITEP}
\newcommand*{\JMU}{James Madison University, Harrisonburg, Virginia 22807, USA}
\newcommand*{\JMUindex}{24}
\affiliation{\JMU}
\newcommand*{\KNU}{Kyungpook National University, Daegu 702-701, Republic of Korea}
\newcommand*{\KNUindex}{25}
\affiliation{\KNU}
\newcommand*{\LPSC}{LPSC, Universite Joseph Fourier, CNRS/IN2P3, INPG, Grenoble, France
}
\newcommand*{\LPSCindex}{26}
\affiliation{\LPSC}
\newcommand*{\UNH}{University of New Hampshire, Durham, New Hampshire 03824-3568, USA}
\newcommand*{\UNHindex}{27}
\affiliation{\UNH}
\newcommand*{\NSU}{Norfolk State University, Norfolk, Virginia 23504, USA}
\newcommand*{\NSUindex}{28}
\affiliation{\NSU}
\newcommand*{\OHIOU}{Ohio University, Athens, Ohio  45701, USA}
\newcommand*{\OHIOUindex}{29}
\affiliation{\OHIOU}
\newcommand*{\ODU}{Old Dominion University, Norfolk, Virginia 23529, USA}
\newcommand*{\ODUindex}{30}
\affiliation{\ODU}
\newcommand*{\RPI}{Rensselaer Polytechnic Institute, Troy, New York 12180-3590, USA}
\newcommand*{\RPIindex}{31}
\affiliation{\RPI}
\newcommand*{\URICH}{University of Richmond, Richmond, Virginia 23173, USA}
\newcommand*{\URICHindex}{32}
\affiliation{\URICH}
\newcommand*{\ROMAII}{Universita' di Roma Tor Vergata, 00133 Rome Italy}
\newcommand*{\ROMAIIindex}{33}
\affiliation{\ROMAII}
\newcommand*{\MSU}{Skobeltsyn Nuclear Physics Institute, Skobeltsyn Nuclear Physics Institute, 119899 Moscow, Russia}
\newcommand*{\MSUindex}{34}
\affiliation{\MSU}
\newcommand*{\UNIONC}{Union College, Schenectady, NY 12308, USA}
\newcommand*{\UNIONCindex}{35}
\affiliation{\UNIONC}
\newcommand*{\UTFSM}{Universidad T\'{e}cnica Federico Santa Mar\'{i}a, Casilla 110-V Valpara\'{i}so, Chile}
\newcommand*{\UTFSMindex}{36}
\affiliation{\UTFSM}
\newcommand*{\GLASGOW}{University of Glasgow, Glasgow G12 8QQ, United Kingdom}
\newcommand*{\GLASGOWindex}{37}
\affiliation{\GLASGOW}
\newcommand*{\VIRGINIA}{University of Virginia, Charlottesville, Virginia 22901, USA}
\newcommand*{\VIRGINIAindex}{38}
\affiliation{\VIRGINIA}
\newcommand*{\WM}{College of William and Mary, Williamsburg, Virginia 23187-8795, USA}
\newcommand*{\WMindex}{39}
\affiliation{\WM}
\newcommand*{\YEREVAN}{Yerevan Physics Institute, 375036 Yerevan, Armenia}
\newcommand*{\YEREVANindex}{40}
\affiliation{\YEREVAN}

\newcommand*{\NOWCUA}{Catholic University of America, Washington, D.C. 20064}
\newcommand*{\NOWLANL}{Los Alamos National Laborotory, New Mexico, NM}
\newcommand*{\NOWJLAB}{Thomas Jefferson National Accelerator Facility, Newport News, Virginia 23606}
\newcommand*{\NOWCNU}{Christopher Newport University, Newport News, Virginia 23606}
\newcommand*{\NOWGWU}{The George Washington University, Washington, DC 20052}
\newcommand*{\NOWWM}{College of William and Mary, Williamsburg, Virginia 23187-8795}

\author {M.H.~Wood} 
\affiliation{\CANISIUS}
\affiliation{\SCAROLINA}
\author {R.~Nasseripour} 
\affiliation{\GWU}
\author {M.~Paolone} 
\affiliation{\SCAROLINA}
\author {C.~Djalali} 
\affiliation{\SCAROLINA}
\author {D.P.~Weygand} 
\affiliation{\JLAB}
\author {K.P.~Adhikari} 
\affiliation{\ODU}
\author {M.~Anghinolfi} 
\affiliation{\INFNGE}
\author {J.~Ball} 
\affiliation{\SACLAY}
\author {M.~Battaglieri} 
\affiliation{\INFNGE}
\author {V.~Batourine} 
\affiliation{\JLAB}
% ...

\collaboration{The CLAS Collaboration}
\noaffiliation

\date{\today}
\begin{abstract}
Due to their long lifetimes, the $\omega$ and $\phi$ mesons are the ideal candidates for the study of possible modifications of the in-medium meson-nucleon interaction through their absorption inside the nucleus.  During the E01-112 experiment at the Thomas Jefferson National Accelerator Facility, the mesons were photoproduced from $\dnuc$, $\cnuc$, $\tinuc$, $\fenuc$, and $\pbnuc$ targets.  This paper reports the first measurement of the ratio of nuclear transparencies for the $\pair$ channel.  The ratios indicate larger in-medium widths compared with what have been reported in other reaction channels.
\end{abstract}


\maketitle

The properties of hadrons, such as their masses and widths, are predicted to be modified in dense and/or hot nuclear matter. Particular attention has been given to the modifications of the properties of vector mesons~\cite{rapp00,post04,rev-mosel,accardi} in ordinary nuclear matter where chiral symmetry is predicted to be partially restored. Different models predict relatively large measurable changes in the mass and/or the width of these mesons~\cite{brown,hatsuda,bernard1988,rapp}. The $\pair$ decay channel of the $\rho$, $\omega$, and $\phi$ has negligible final-state interactions (FSI), providing an ideal tool to study the possible in-medium modifications of these mesons. Due to its short lifetime, the $\rho$ meson has a substantial probability of decaying in the nucleus, while the $\omega$ and $\phi$ mesons tend to decay outside.  In order to directly observe any in-medium modifications of the $\omega$ and $\phi$ mesons, one has to select those produced with very small momentum relative to the nucleus, so that they remain in the nucleus until they decay.  With such an experiment, it is difficult to collect enough statistics in a reasonable amount of time.

The E01-112 experiment at the Thomas Jefferson National Accelerator Facility (JLab) has studied possible modifications of the $\rho$, $\omega$, and $\phi$ mesons with momenta greater than 0.8~GeV.  The results for the $\rho$ meson have been reported in Refs.~\cite{g7a-prl,g7a-prc}.  In this paper, we present the results for the $\omega$ and $\phi$ mesons.  The observed $\omega$ and $\phi$ mesons in this experiment are produced at relatively large momenta, with a majority of them decaying outside of the nucleus.  However, their in-medium width can be extracted from their absorption in the nuclei.  The in-medium width is given by $\Gamma = \Gamma_{0} + \Gamma_{coll}$, where  $\Gamma_{0}$ is the natural width in vacuum and $\Gamma_{coll}$ is the width due to collisional broadening.  Using the low-density theorem~\cite{dover71}, $\Gamma_{coll} = \gamma \rho v \sigma^{*}_{VN}$, where $\gamma$ is the Lorentz factor, $\rho$ is the mean effective nuclear density, $v$ is the velocity of the meson, and $\sigma^{*}_{VN}$ is the meson-nucleon total cross section in the nucleus.  The low-density theorem assumes the density of nucleons is small so that the meson scatters off of a single nucleon.  Thus, meson modifications can be extracted from a measurement of the in-medium meson-nucleon cross section.

A broadening of the in-medium width can be attributed to a number of effects.  A lowering of the meson mass raises the number of sub-threshold channels available to the meson.  Another effect is the modification of their hadronic components.  In the case of the $\omega$ meson, one can consider it as a virtual $\pi \rho$ pair, where both the pion and the $\rho$ meson are modified by the medium.  Similarly, the $\phi$ meson can be considered a virtual $\Kpair$ pair, with in-medium modification of the kaons.  These medium effects are discussed in detail by Refs.~\cite{mue_omega,kuskalov1,kuskalov2,titov} for the $\omega$ meson and by Refs.~\cite{mue_phi,oset01,cabrera03,cabrera04,magas05} for the $\phi$ meson.  

An extensive review of previous experiments searching for modifications of the light vector mesons can be found in Ref.~\cite{rev-mosel}.  For the $\omega$ meson, the  KEK-E325 and CB-ELSA/TAPS experiments disagree on the measured masses and widths.  In the KEK-E325 experiment, the $\rho$, $\omega$, and $\phi$ mesons were produced with a 12-GeV proton beam on $\cnuc$ and $\cunuc$ and were detected by  their decays to $\pair$.  The CB-ELSA/TAPS collaboration investigated the reaction $\gamma A \rightarrow \omega X \rightarrow \pi^{0} \gamma X$ with targets of deuterium and $\nbnuc$.  In this measurement, the $\rho \rightarrow \pi^{0} \gamma$ decay channel is suppressed by two orders of magnitude in comparison to the $\omega$ meson; however, the outgoing $\pi^{0}$'s interact strongly in the nucleus.  The KEK-E325 collaboration reported a 9\% decrease in the $\omega$-meson mass with no change in the width~\cite{kek-new}.  The CB-ELSA/TAPS results indicate no mass shift~\cite{nanova} and an in-medium width of 130-150~MeV~\cite{kotulla}.  

For the $\phi$ meson, the KEK-E325 collaboration reported no change to the mass, but an in-medium width of 14.4~MeV~\cite{sakuma,muto}.  This disagrees with the measurement at SPring8.  At the SPring8 facility, the in-medium $\phi N$ interaction was studied with the reaction $\gamma A \rightarrow \phi X \rightarrow K^{+} K^{-} X$ on $\linuc$, $\cnuc$, $\alnuc$, and $\cunuc$ targets~\cite{ishikawa}.  This measurement had the complication of kaon final-state interactions.  They reported the in-medium $\phi N$ cross section to be $35^{+17}_{-11}$~mb, which corresponds to a collisionally-broadened width of 35~MeV.  The SPring8 result is in agreement with two other JLab measurements: one is the coherent $\phi$-meson photoproduction on the deuteron~\cite{mibe} and the other is from the ${\rm d(\gamma,pK^{+}K^{-})n}$ reaction~\cite{qian}.  This paper addresses these discrepancies with new measurements from JLab of the in-medium widths of both mesons in heavy nuclei.

For this JLab experiment, we studied the in-medium meson widths with the reaction $\gamma A \rightarrow V X \rightarrow e^{+} e^{-} X$, where $V$ is either the $\omega$ or $\phi$ meson.  This measurement is the only one with electromagnetic probes in both the incoming and outgoing channels.  A beam of real photons illuminated the entire nuclear volume, while the lepton pair suppressed FSI effects.  The experiment was conducted with the CEBAF Large Acceptance Spectrometer (CLAS)~\cite{clas} and the photon tagging facility~\cite{tagnim} in Hall B.  The CLAS detector is ideal because of its designed multi-particle detection capabilities and its excellent electron identification.  The photon beam energy ranged up to 4~GeV.  The targets were $\dnuc$, $\cnuc$, $\tinuc$, $\fenuc$, and $\pbnuc$.  Since the $\tinuc$ and $\fenuc$ nuclei have similar sizes, their data were combined and the notation $\tife$ is used throughout the paper.  All target materials were in the beam simultaneously, separated by 2.0~cm along the beam direction, to eliminate possible multiple scattering of the outgoing leptons.  The target thicknesses were 1~g/${\rm cm^{2}}$ with up to 10\% variations between the individual targets.  More details about the current experiment can be found in Refs.~\cite{g7a-prl,g7a-prc}.

To access the $A$-dependence of the modifications, the nuclear transparency was determined for the individual targets.  The transparency is defined as $T_{A} = \sigma_{A}/A\sigma_{N}$, where $\sigma_{A}$ and $\sigma_{N}$ are the nuclear and nucleon total cross sections, respectively.  To obtain the experimental cross sections, the yields, acceptance, and target thicknesses were determined.  The number of incident photons canceled in the ratio of cross sections.

\begin{figure}[htpb]
\centering
\includegraphics[width=7.0cm]{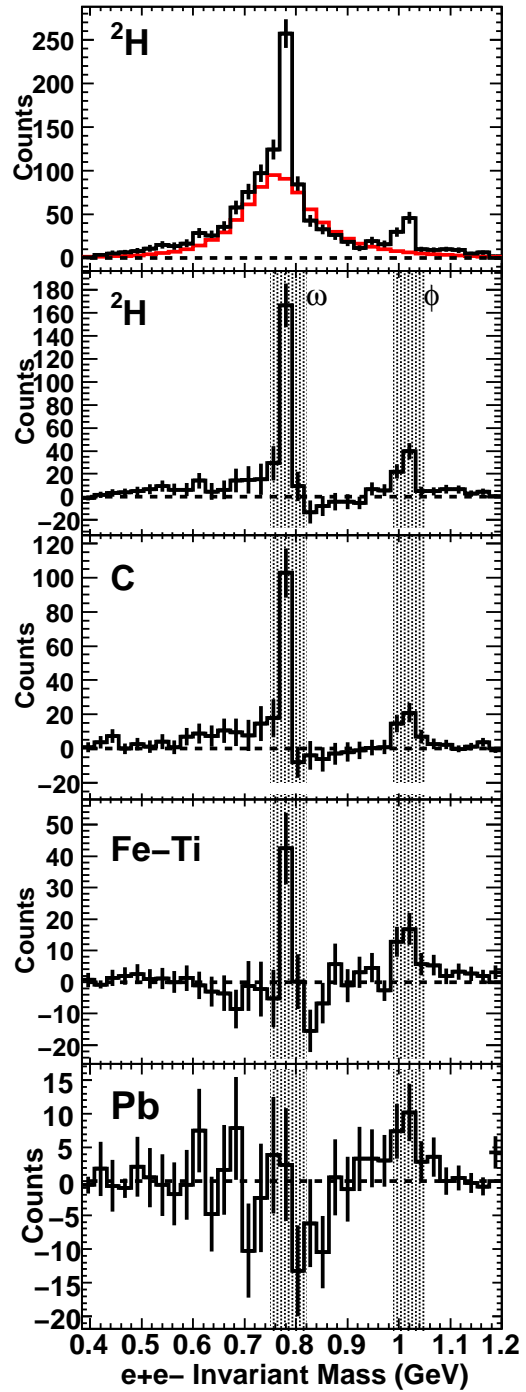}
\caption[]{\small{(color online) The $\pair$ mass spectra before and after the subtraction of the $\rho$-meson contribution.  The top $\dnuc$ plot shows the mass spectrum before subtraction.  The red histogram is the normalized $\rho$-meson line shape.  The remaining histograms are after subtraction and are labeled by target. The shaded bands are the integration ranges of the $\omega$ and $\phi$ mesons.}}
\label{fig:MassSpec}
\end{figure}

To extract the meson yields, a spectrum of the $\pair$ mass was reconstructed for each target, and the $\rho$-meson shape has been subtracted.  The shape of the $\rho$ meson was taken from the realistic GiBUU transport model~\cite{effenberger1,effenberger2}, which reproduced current JLab $\rho$ data~\cite{g7a-prl,g7a-prc}.  The $\rho$-meson contribution was normalized to match the $\pair$ mass spectrum in the region between the $\omega$ and $\phi$ mesons ($0.85 \leq M \leq 0.95$~GeV) and subtracted.  The top plot in Fig.~\ref{fig:MassSpec} shows the mass spectrum for the $\dnuc$ target before the subtraction, along with the normalized $\rho$-meson line shape.  The other plots in the figure show the mass spectra for the four targets after removal of the $\rho$-meson contribution.  These mass spectra differ from those shown in Refs.~\cite{g7a-prl,g7a-prc} because a higher cut of 0.7~GeV was applied to the lepton momentum.  The yields were the sum of the counts in the region of the mesons ($0.75 \leq M_{\omega} \leq 0.82$~GeV, $0.99 \leq M_{\phi} \leq 1.05$~GeV).  The positive counts below and the negative counts above the $\omega$-meson mass may be indicative of $\rho$-$\omega$ interference.  Instead of a fit to extract the amplitudes and phase angle, upper and lower limits on the yields were determined by summing the absolute value of the counts and by summing the positive and negative counts, respectively.

The acceptances were averaged over the mass range of each individual meson and spanned from about 3-5\% between the targets for the $\omega$-meson channel and 8-12\% for the $\phi$-meson channel.

\begin{figure}[ht]
\centering
\includegraphics[width=7.0cm]{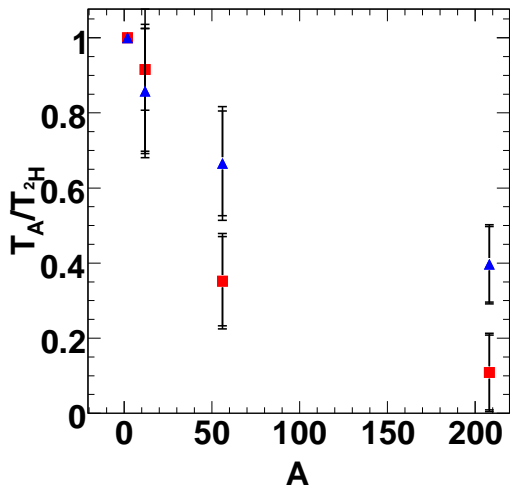}
\caption[]{\small{(color online) Transparency ratios, normalized to deuterium, versus atomic mass for the $\omega$ meson (red squares) and $\phi$ meson (blue triangles).  The inner error bars are the statistical uncertainties.  The outer error bars represent the total uncertainties, statistical and systematic summed in quadrature.  For the $\omega$ meson, the statistical error bars contain the upper and lower limits.}}
\label{fig:TAd2}
\end{figure}

To eliminate the nucleon cross section in the analysis of the transparency, a ratio was made of the heavier target transparencies and the $\dnuc$ transparency, $T_{A}/T_{^{2}H}$.  Fig.~\ref{fig:TAd2} is a plot of the ratios for the $\omega$ and $\phi$ mesons.  For both mesons, the ratio decreases rapidly with $A$, indicating a substantial increase of the in-medium width with target atomic mass.

The systematic uncertainties were determined for the vertex matching of the leptons and the yield extraction.  The leptons in each event were matched by their difference in $z$-position along the beamline, the radial position $r$ in the target, and by the difference in their vertex time $\Delta t$.  More information on these quantities and on the determination of the related systematic uncertainties can be found in Refs.~\cite{g7a-prl,g7a-prc}.  For both meson channels, the range in the uncertainties for vertex matching was from 0.5-7.0\% as a function of $A$, with the $\pbnuc$ target having the highest value.  For the yield extraction, a systematic study of the $\rho$-meson subtraction, varying the bin width and a lepton momentum cut~\cite{g7a-prl,g7a-prc}, produced uncertainties from 10-20\% for the $\omega$-meson analysis and 3-8\% for the $\phi$-meson analysis.  These uncertainties dominated the error.  The total systematic uncertainties were 20\% ($\cnuc$), 11\% ($\tife$), and 18\% ($\pbnuc$) for the $\omega$ meson and 7\% ($\cnuc$), 9\% ($\tife$), and 8\% ($\pbnuc$) for the $\phi$ meson.  It should be noted that the lower limit on the $\omega$-meson transparency ratio for the $\pbnuc$ data is zero.  Therefore, it was only meaningful to apply the systematic uncertainty to the upper limit for this data point.  

\begin{figure}[htpb]
\centering
\includegraphics[width=8.0cm]{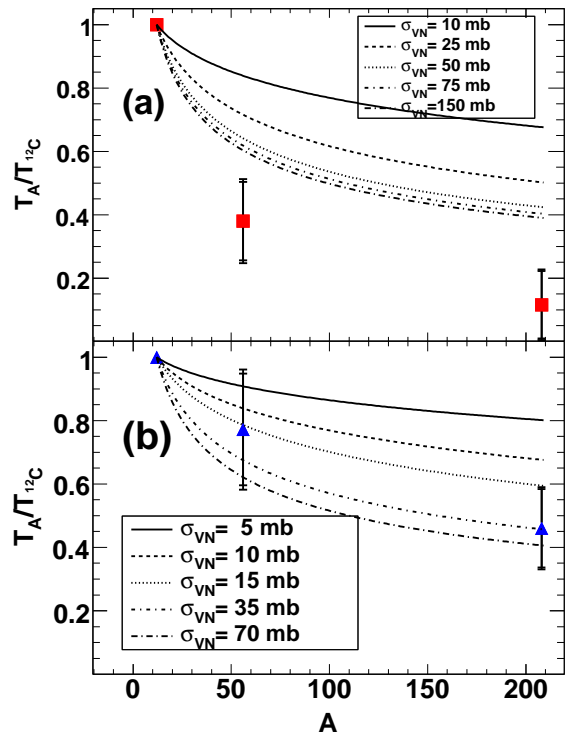}
\caption[]{\small{(color online) Transparency ratios, normalized to carbon, versus atomic mass for the $\omega$ meson (a) and $\phi$ meson (b).  The curves are Glauber calculations where the $\sigma_{VN}$ is given in the legend.  The error bars are described in Fig.~\ref{fig:TAd2}.}}
\label{fig:TAc}
\end{figure}

In order to extract information about the in-medium $\omega N$ and $\phi N$ cross sections, a Glauber calculation was made following the prescription of Refs.~\cite{mue_omega,bauer}.  Fig.~\ref{fig:TAc} shows a comparison of the data and the Glauber calculations with a range of in-medium cross sections.  For the $\phi$ meson, $\sigma_{VN}$ is in the range of 16-70~mb.  The transparency ratios were normalized to the carbon data ($T_{A}/T_{C}$) in order to compare with previous data and calculations in Ref.~\cite{kotulla}.  

\begin{figure}[htpb]
\centering
\includegraphics[width=8.0cm]{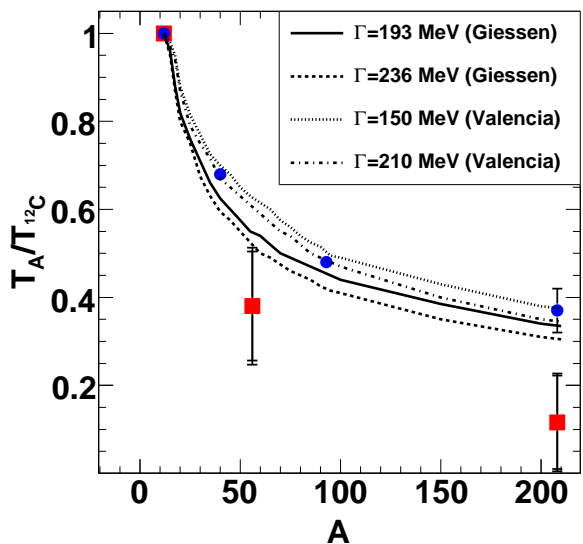}
\caption[]{\small{(color online) Comparisons of the $\omega$-meson transparency ratios, normalized to carbon, with theoretical calculations and previous experimental results.  The red squares are the JLab data, while the blue circles are from the CB-ELSA/TAPS experiment~\cite{kotulla}.  The curves are the calculations from the Giessen and Valencia groups.  The legend lists the in-medium widths employed in each calculation.  The JLab error bars are described in Fig.~\ref{fig:TAd2}.}}
\label{fig:TAcomp}
\end{figure}

The $\omega$ meson data are not reproduced by the Glauber calculations.  To investigate further, these transparency ratios were compared with the results by the CB-ELSA/TAPS experiment and the calculations by the Valencia and Giessen groups~\cite{kotulla}.  The calculations by the Valencia group are based on photoproduction of the $\omega$ meson off the nucleon and a Monte Carlo simulation of the propagation of the meson and its decay products through the nucleus~\cite{kuskalov1}.  The Giessen model also starts with elementary production; however, a Boltzmann-Uehling-Uhlenbeck transport calculation is used for the propagation~\cite{mue_omega}.  The comparisons are shown in Fig.~\ref{fig:TAcomp}.  The present results show much greater broadening than the theoretical calculations and the CB-ELSA/TAPS results.  Even with the upper limit placed on the $\pbnuc$ data point, the in-medium width is greater than 200~MeV.

The in-medium widths of the $\omega$ and $\phi$ mesons have been extracted by utilizing the ratio of nuclear transparencies.  Large absorption strengths for both mesons have been observed, which implies an increase of the in-medium widths.  For the $\phi$ meson with its strange-quark content, the modifications are expected to originate from the dressing of the virtual $K\bar{K}$ loops and not from nucleon resonances coupling to the $\phi N$ system~\cite{mue_phi}.  The width is predicted to increase with very little change in the mass~\cite{oset01,cabrera03}.  The experimental results show an increase but disagree on the size.  The KEK-E325 and SPring8 collaborations measured widths of 14.4~MeV~\cite{kek-new,sakuma,muto} and 35~MeV~\cite{ishikawa}, respectively.  Our result is 23-100~MeV and overlaps with the SPring8 measurement.  In calculating the widths, $\rho = 0.5 \rho_{0} = 0.08~{\rm fm}^{-3}$ is used for the effective density.  These results imply that the in-medium $\phi N$ total cross section is much larger than in free-space and is comparable with other meson-nucleon total cross sections ($\sigma_{\rho N}$, $\sigma_{\omega N}$, and $\sigma_{\eta N}$) in free space~\cite{bauer}.  Ref.~\cite{qian} offers an interesting explanation for the larger $\phi N$ cross section in deuterium in their data and that of Ref.~\cite{mibe}.  They claim the increase is related to the mixing between the $\omega$ and $\phi$ mesons.  The $\omega$ meson is produced on one nucleon and elastically scatters off of a second nucleon.  The matrix element of the $\omega$ meson has the same structure as the $\phi$-meson rescattering matrix element, so there is an enhancement to the effective $\sigma_{\phi N}$.  An extension of our experiment with more statistics can test this idea further, since both mesons are produced in nuclei and are present in the reconstructed mass spectrum.~\cite{g7b-prop}.

For the $\omega$ meson, our results indicate more absorption than can be accounted for by the published CB-ELSA/TAPS results~\cite{kotulla}.  Unlike our measurement, the CB-ELSA/TAPS experiment has to contend with the final-state interactions of the $\pi^{0}$.  In addition, the disagreement may be due to the average meson-momenta in each measurement, 1.1~GeV (CB-ELSA/TAPS) and 1.7~GeV (JLab).  To perform a proper comparison, each result needs to be scaled to match the kinematics.  In Ref.~\cite{kotulla}, the $\pbnuc$ data is divided into 5 momentum bins with the highest bin at about 1.7~GeV.  The transparency ratio for this data point is about twice as large as our $\pbnuc$ data point in Fig.~\ref{fig:TAcomp}.  Even with similar meson momentum, there is still a discrepancy.  Our larger absorption may be evidence of destructive $\rho$-$\omega$ interference in the medium.  Since the CB-ELSA/TAPS decay channel is $\omega \rightarrow \pi^{0} \gamma$, the mixing with the $\rho$ meson is negligible.  With our experiment, both mesons decay into $\pair$ with similar strengths.  The E04-005 experiment at JLab collected data in 2008 with a high-intensity photon beam on a $\hnuc$ target~\cite{g12-prop}.  These data are being analyzed for the elementary processes, that will provide necessary information for the in-medium interpretation.  The results will be published in a forthcoming article.

The stronger $\omega$-meson absorption cannot be explained with current theoretical calculations.  Our Glauber calculation cannot reproduce the data, even as it converges at high cross sections.  Moreover, both the Giessen and Valencia models~\cite{kotulla} indicate that the in-medium width should be greater than 200~MeV.

As with the $\omega$ meson, the $\phi$-meson experiments had different average meson momenta.  The values are $<$1.25~GeV (KEK-E325), 1.7~GeV (SPring8), and 2~GeV (JLab).  Taken together, the results show evidence of a momentum dependence of the in-medium width.  The E08-018 experiment at JLab will allow for a study of the momentum dependence and will produce more definitive results of the meson absorption~\cite{g7b-prop}.  

Our results are the first to measure the $\omega$- and $\phi$-meson absorptions by means of their rare decay into $\pair$ and complement previous measurements with hadronic decay channels.  These results will impact relativistic heavy-ion studies and medium modification searches.  For the relativistic heavy-ion transport calculations, they provide more realistic values of the in-medium widths.  For direct measurements of the medium-modified mass and width with low-momentum mesons, there is a reduction in the sensitivity since these mesons are more likely to interact in the medium before they decay. 

Our results show a strong absorption for both mesons.  The $\phi$-meson result indicates that the in-medium cross section is much larger than that in free space.  The free space value is less than 10~mb~\cite{lipkin,behrend}, while our measurement is in the range of 16-70~mb.  This range is closer to the free-space $\rho N$ and $\omega N$ cross sections.  The $\omega$-meson result shows greater absorption than the CB-ELSA/TAPS measurement.  Our transparency ratio for the $\pbnuc$ nucleus is around 14\%.  The theoretical calculations cannot account for the increased absorption.  Other effects such as momentum dependence or in-medium $\rho$-$\omega$ interference may be responsible.

We would like to thank the staff of the Accelerator and Physics Divisions at Jefferson Laboratory who made this experiment possible. This work was supported in part by the U.S. Department of Energy, the National Science Foundation, the Research Corporation, the Italian Istituto Nazionale di Fisica Nucleare, the French Centre National de la Recherche Scientifique, the French Commissariat \'a l'Energie Atomique, the National Research Foundation of Korea, the U.K. Science and Technology Facilities Council (STFC), Deutsche Forschungsgemeinschaft, and the Chilean Comisi\'on Nacional de Investigaci\'on Cient\'ifica y Tecnol\'ogica (CONICYT).  Jefferson Science Associates (JSA) operates the Thomas Jefferson National Accelerator Facility for the United States Department of Energy under contract DE-AC05-84ER40150. The authors appreciate the theoretical support provided by U.~Mosel and P.~Muehlich.

\end{document}